\documentstyle{article}
\begin{document}
\vspace*{1.1cm} \centerline{\large\bf BITWISTOR FORMULATION OF MASSIVE SPINNING PARTICLE} \vspace{0.5cm} \centerline{\large {\bf
S. Fedoruk}$^1$ {\large {\bf and}} {\bf V.G. Zima}$^2$} \centerline{$^1$ {\small\it Ukrainian Engineering--Pedagogical Academy,
61003 Kharkiv, 16 Universitetska Str., Ukraine}} \centerline{\small\it e-mail: fed@postmaster.co.uk} \centerline{$^2$ {\small\it
Kharkiv National University,
     61077 Kharkiv, 4 Svobody Sq., Ukraine}}
\centerline{\small\it e-mail: zima@pht.univer.kharkov.ua,
zima@postmaster.co.uk} \vspace{1.3cm} {\small \noindent Twistor
formulation of massive arbitrary spin particle has been
constructed. Twistor space of such particle is formed two twistors
and two complex scalars which form together `bosonic
supertwistor'. The formulation is deduced from space--time one for
spinning particle by means of introducing auxiliary harmonic
variables and consequent partial fixing of gauges. It is carried
out the canonical quantization of twistor massive particle with
nonzero spin. It is found the eigenvalues of Casimir operators on
particle states and harmonic expansion of wave function in
spectrum.\\
{\bf KEYWORDS:} twistors; spinning particle; harmonic variables.}
\par\bigskip
At present the twistor formulations of point-like and extended
objects are acquired grate role in modern particle physics~[1-16].
Up to now massless (super)particles was considered in
(super)twistor approach in the main. In this article we construct
twistor formulation of massive spinning particle.

At construction of twistor formulation of massive spinning
particle it is necessary to solve the question of describing of
massive spinning states in form of twistors. The natural method in
resolving this task is introducing greater than one
twistors~[1-4], [16] i. e. in some terminology it is obtained the
description of massive state in form of two or more massless
states~[8]. But the problem of spin description and in fact the
right chooses of variables in twistor formulation, the constraints
and Lagrangian are remained not solved finally.

The constructive way to finding of twistor formulation of spinning
particle implies using the appropriate space--time formulation.
For this aim from all space--time formulations the more
appropriate formulation are those in which the spin degrees of
freedom are described by means of commuting variables. Also from
such formulations there are appropriate ones in which spin
variables are spinors (for obtaining arbitrary spins including
half--integer ones) and describing of arbitrary spins is realized
in uniform way. The formulation relativistic spinning particle
with index spinor~[17-20] is more appropriate formulation for
these aims.
\par\bigskip
\centerline{\bf SPINNING PARTICLE IN INDEX SPINOR FORMULATION}
\par
From all space--time formulations of spinning particle the
formulations with spinning variables of bosonic type are
appropriate ones to construction of twistor formulation of it. For
this end we prefer formulation of spinning particle with index
spinor. Its advantages are confined in use spinor variables for
description of particle spin that improve transition to twistor
formulation. Also spinning particle with index spinor has a some
analogy with usual superparticle which use Grassmannian spinor
variable. Therefore for our aim we can exploit the some elements
of transition from space--time formulation of superparticle to
twistor one.

In index spinor formalism spinning particle is described with
space--time vector $x^\mu$ and commuting Weyl spinor
$\zeta^\alpha$. In first order formalism its Lagrangian has the
form~[17-20]
\begin{equation}\label{1}
L=p\dot\Pi-V(p^2+m^2)-\Lambda(\zeta\hat p\bar\zeta-j)\,,
\end{equation}
where the bosonic `superform' is
$$
\Pi\equiv\dot\Pi\,d\tau=dx-id\zeta\sigma\bar\zeta+i\zeta\sigma
d\bar\zeta\,.
$$
Here $p_\mu$ is momentum vector of particle with mass $m$. Real
scalars $V$ and $\Lambda$ are Lagrange multipliers. In spinor
notation
\begin{equation}\label{lagr-s,m}
L= -{\textstyle \frac{1}{2}}
p_{\alpha\dot\alpha}\dot\Pi^{\dot\alpha\alpha} + {\textstyle
\frac{1}{2}} V (p_{\alpha \dot\alpha}p^{\dot\alpha\alpha}-2m^2) -
\Lambda (\zeta^\alpha p_{\alpha \dot\alpha}\bar\zeta^{\dot\alpha}
-j)\, ,
\end{equation}
where the bosonic `superform' is
$$
\Pi^{\dot\alpha\alpha}\equiv\dot\Pi^{\dot\alpha\alpha}d\tau \equiv
dx^{\dot\alpha\alpha} +i\bar\zeta^{\dot\alpha}d\zeta^\alpha -
id\bar\zeta^{\dot\alpha}\zeta^\alpha \, .
$$
We use spinor notations which coincide with~[21]. In particular,
$p_{\alpha\dot{\alpha}}=p_\mu \sigma^\mu_{\alpha\dot{\alpha}}$,
$x^{\dot{\alpha}\alpha}=x^\mu \sigma_\mu^{\dot{\alpha}\alpha}$,
where matrices $\sigma_\mu$ satisfy
$\sigma^{\dot{\alpha}\alpha}_{(\mu}\sigma_{\nu)\alpha\dot{\beta}}=
-\eta_{\mu\nu}\delta^{\dot{\alpha}}_{\dot{\beta}}$ with
$\eta_{\mu\nu}=\textrm{diag}(-,+,+,+)$.

Apart from the constraints inserted into the action explicitly, i.
e. the mass constraint
\begin{equation}\label{s1}
T\equiv{\textstyle \frac{1}{2}}(p^2 +m^2) \approx 0 \, ,
\end{equation}
and the spin one
\begin{equation}\label{s2}
\zeta^\alpha p_{\alpha\dot\alpha}\bar\zeta^{\dot\alpha}-j\approx 0
\end{equation}
the Hamiltonization~[8] of the theory reveals also the spinor
Bose--constraints
\begin{equation}\label{s3}
d_{\zeta\alpha}\equiv ip_{\zeta\alpha}
+p_{\alpha\dot\alpha}\bar\zeta^{\dot\alpha}\approx 0\, ,\qquad
\bar d_{\zeta\dot\alpha}\equiv -i\bar p_{\zeta\dot\alpha}
+\zeta^\alpha p_{\alpha\dot\alpha}\approx 0\, .
\end{equation}
On the constraints surface the spin constraint is equivalent to
the constraint
\begin{equation}\label{s4}
\emph{\textbf{S}}\equiv S-j\equiv{\textstyle
\frac{i}{2}}(\zeta^\alpha p_{\zeta\alpha}-\bar
p_{\zeta\dot\alpha}\bar\zeta^{\dot\alpha})-j\approx 0\, ,
\end{equation}
because $\emph{\textbf{S}}\equiv\frac{1}{2}(\zeta d_\zeta-\bar
d_\zeta\bar\zeta)+(\zeta\hat p\bar\zeta-j)$.

Immediately it is found the constraint algebra, whose nontrivial
brackets are
$$
\{ d_\zeta,\bar d_\zeta\}=2i\hat p\, ,\qquad \{
\emph{\textbf{S}},d_\zeta\}={\textstyle \frac{i}{2}}\,d_\zeta\, ,
\qquad \{\emph{\textbf{S}},\bar d_\zeta\}=-{\textstyle
\frac{i}{2}}\,\bar d_\zeta\, .
$$
So, the constraints $T$ and $S$ belong to the first class whereas
the spinor constraints $d_{\zeta\alpha}$ and $\bar
d_{\zeta\dot\alpha}$ relate to the second class for particle with
nonzero mass, i. e. $\hat p\tilde p=m^2>0$. Certainly in the
procedure of Hamiltonization the spinor constraints are primary
whereas the mass constraints and the spin one are constraints of
the second step of the procedure. As a consequence of
reparametrization invariance the total Hamiltonian is a linear
combination of the first class constraints $T\approx 0$ (\ref{s1})
and $\emph{\textbf{S}}\approx 0$ (\ref{s4}).

After quantization the wave function of the spinning particle is
expressed by (anti)\-ho\-lo\-mor\-phic polynomial on spinor
$\zeta$~[17]. Therefore spinor $\zeta$ was been called index
spinor. The information about Lorentz properties of wave function
is encoded in a polynomial structure on index spinor. The spin of
particle after quantization (only one value of spin!) is equal
constant $j$ in Lagrangian renormalized by ordering constants.
\par\bigskip
\centerline{\bf TWISTOR FORMULATION FROM LORENTZ HARMONIC
APPROACH}
\par
Let us obtain the twistor formulation from space--time
formulation~(\ref{lagr-s,m}).

We introduce two spinors
\begin{equation}\label{lambda2}
v_\alpha^i \, ,\quad \bar v_{\dot\alpha i} =
\overline{(v_\alpha^i)}  \, ,\quad i=1,2
\end{equation}
which considered as Lorentz harmonics~[22-24]. Its formed $2\times
2$ complex matrix with unit determinant. The harmonic spinors
$v_\alpha^i$ are subjected to the conditions
\begin{equation}\label{norm-harm}
h\equiv v^{\alpha i}v_{\alpha i} +2\approx 0 \, ,\quad \bar
h\equiv\bar v_{\dot\alpha i} \bar v^{\dot\alpha i} +2\approx 0\, .
\end{equation}
where $v_{\alpha i} =\epsilon_{ij}v_\alpha^j$, $\bar
v_{\dot\alpha}^i =\epsilon^{ij}\bar v_{\dot\alpha j}$ and
components of skew--symmetric tensor $\epsilon^{ij}$ are equal
matrix elements of matrix $i\sigma_2$,
$\epsilon^{ij}\epsilon_{jk}=\delta^i_k$. The
conditions~(\ref{norm-harm}) can be inscribed in the equivalent
form
$$
v^{\alpha i}v_{\alpha}^j -\epsilon^{ij}\approx 0 \, ,\quad \bar
v_{\dot\alpha i} \bar v^{\dot\alpha}_j -\epsilon_{ij}\approx 0
$$
or also in form
$$
v^{\alpha i}v^{\beta}_i -\epsilon^{\alpha\beta}\approx 0 \, ,\quad
\bar v_{\dot\alpha i} \bar v_{\dot\beta}^i
-\epsilon_{\dot\alpha\dot\beta}\approx 0 \, .
$$

Let us complement the system~(\ref{lagr-s,m}) by pure gauge sector
of Lorentz harmonics. Namely we add to Lagrangian~(\ref{lagr-s,m})
standard kinetic terms for harmonics $v_\alpha^i$, $\bar
v_{\dot\alpha i}$ and canonically conjugate variables
$p_v{}^\alpha_{i}$, $\bar p_v{}^{\dot\alpha i}$, $\{v_\alpha^i,
p_v{}^{\beta}_{j} \}= \delta^\beta_\alpha \delta_j^i$, $\{\bar
v_{\dot\alpha i}, \bar p_v{}^{\dot\beta j} \}=
\delta^{\dot\beta}_{\dot\alpha}\delta^j_i$  and linear combination
of the full set of the constraints, the coefficients of which are
Lagrange multipliers. The number of constraints must be sufficient
to exclude all harmonic variables. In addition to kinematic
constraints~(\ref{norm-harm})  we impose the following natural set
of constraints on harmonic variables
\begin{equation}\label{const-om}
p_v{}^{\alpha}_{i}\approx 0 \, ,\qquad \bar p_v{}^{\dot\alpha
i}\approx 0 \, ,
\end{equation}
i.e. all conjugate variables for harmonics  $v_\alpha^i$, $\bar v_{\dot\alpha i}$ are zero in weak sense. Of course, the
constraints~(\ref{const-om}) mean that the variables $v_\alpha^i$, $\bar v_{\dot\alpha i}$ are constant on equations of motion,
$\dot v_\alpha^i =0$, $\dot{\bar v}_{\dot\alpha i} =0$.

The system of constraints~(\ref{norm-harm}) and (\ref{const-om})
contain two pairs of second class constraints and six of first
class constraints. The separation of constraints~(\ref{const-om})
on classes is realized by projection of them on spinors
$v_\alpha^i$, $\bar v_{\dot\alpha i}$. Because of nonsingularity
of harmonic matrix $v_\alpha^i$ (\ref{norm-harm}), the set of
constraints~(\ref{const-om}) and Lorentz--invariant constraints
\begin{equation}\label{const-om-pr}
p_v{}^{\alpha}_{i}v_\alpha^j\approx 0 \, ,\qquad \bar
v_{\dot\alpha i}\bar p_v{}^{\dot\alpha j}\approx 0
\end{equation}
are equivalent.

The trace parts of constraints~(\ref{const-om-pr})
\begin{equation}\label{const-om-tr}
p_v{}^{\alpha}_{i}v_\alpha^i\approx 0 \, ,\qquad \bar
v_{\dot\alpha i}\bar p_v{}^{\dot\alpha i}\approx 0
\end{equation}
are conjugate ones for kinematic constraints~(\ref{norm-harm}). In
real quantities the constraints
\begin{equation}\label{norm-h}
i(h-\bar h)=i(v^{\alpha i}v_{\alpha i} - \bar v_{\dot\alpha i}
\bar v^{\dot\alpha i})\approx 0 \, ,\quad h+\bar h=v^{\alpha
i}v_{\alpha i} + \bar v_{\dot\alpha i} \bar v^{\dot\alpha i}
+4\approx 0
\end{equation}
and
\begin{equation}\label{const-Tr}
{\cal D}_0 \equiv i(p_v{}^{\alpha}_{i}v_\alpha^i - \bar
v_{\dot\alpha i}\bar p_v{}^{\dot\alpha i})\approx 0 \, ,\quad
{\cal B}_0 \equiv p_v{}^{\alpha}_{i}v_\alpha^i + \bar
v_{\dot\alpha i}\bar p_v{}^{\dot\alpha i}\approx 0
\end{equation}
form pairs $(i(h-\bar h),{\cal D}_0)$ and $(h+\bar h,{\cal B}_0)$
of conjugate each other second class constraints.

The traceless parts of constraints~(\ref{const-om-pr}), which
commute with constraints~(\ref{norm-h}) and (\ref{const-Tr}), are
first class constraints. It is convenient to represent these
constraints in form real Lorentz invariant $3$--vectors
\begin{equation}\label{const-Vec}
{\cal D}_r \equiv {\textstyle \frac{i}{2}}(\sigma_r )_i{}^j
(p_v{}^{\alpha}_{j}v_\alpha^i - \bar v_{\dot\alpha j}\bar
p_v{}^{\dot\alpha i})\approx 0 \, ,\quad {\cal B}_r \equiv
{\textstyle \frac{1}{2}}(\sigma_r )_i{}^j
(p_v{}^{\alpha}_{j}v_\alpha^i + \bar v_{\dot\alpha j}\bar
p_v{}^{\dot\alpha i})\approx 0
\end{equation}
where matrices $\sigma_r$, $r=1,2,3$ are usual Hermitian Pauli
matrices.

Thus spinning particle, added pure gauge sector of Lorentz
harmonics, is described phase space variables $x^\mu$, $p_\mu$;
$\zeta^\alpha$, $\bar\zeta^{\dot\alpha}$, $p_{\zeta\alpha}$, $\bar
p_{\zeta\dot\alpha}$; $v_\alpha^i$, $\bar v_{\dot\alpha i}$,
$p_v{}^\alpha_{i}$, $\bar p_v{}^{\dot\alpha i}$ and
constraints~(\ref{s1}), (\ref{s3}), (\ref{s4}), (\ref{norm-harm}),
(\ref{const-Tr}), (\ref{const-Vec}). Let us exclude with using of
part of the constraints the variables $x$, $p$, $\zeta$, $p_\zeta$
and c.c. which are used in space--time formulation.

For that it is convenient to transform by Lorentz harmonics the
initial variables $x$, $p$, $\zeta$, $p_\zeta$ and c.c. to
Lorentz--invariant quantities
\begin{equation}\label{x}
x^{(0)} ={\textstyle \frac{1}{2}}\bar v_{\dot\alpha k}
x^{\dot\alpha\alpha}v_{\alpha}^k\, , \quad x_{(r)} ={\textstyle
\frac{1}{2}}\bar v_{\dot\alpha j}
x^{\dot\alpha\alpha}v_{\alpha}^i(\sigma_r )_i{}^j\, ,
\end{equation}
\begin{equation}\label{p}
p_{(0)} ={\textstyle \frac{1}{2}}v^{\alpha}_k
p_{\alpha\dot\alpha}\bar v^{\dot\alpha k}\, , \quad p_{(r)}
={\textstyle \frac{1}{2}}v^{\alpha}_j p_{\alpha\dot\alpha}\bar
v^{\dot\alpha i}(\sigma_r )_i{}^j\, ,
\end{equation}
\begin{equation}\label{k}
\xi^i =m^{1/2}\zeta^\alpha v_\alpha^i \, ,\quad \bar\xi_i
=m^{1/2}\bar v_{\dot\alpha i}\bar\zeta^{\dot\alpha} \, ,
\end{equation}
\begin{equation}\label{pk}
p_{\xi i}= m^{-1/2}v^{\alpha}_i p_{\zeta\alpha}\, ,\quad \bar
p_{\xi}^i= -m^{-1/2}\bar p_{\zeta\dot\alpha} \bar v^{\dot\alpha i}
\, .
\end{equation}

At this transition harmonic variables $v_\alpha^i$, $\bar
v_{\dot\alpha i}$, $p^\alpha_{vi}$, $\bar p_v^{\dot\alpha i}$
transform to variables
\begin{equation}\label{l}
\lambda_\alpha^i =m^{1/2} v_\alpha^i \, ,\quad
\bar\lambda_{\dot\alpha i} =m^{1/2}\bar v_{\dot\alpha i} \, .
\end{equation}
Momenta $\bar\omega^\alpha_i$, $\omega^{\dot\alpha i}$ for
$\lambda_\alpha^i$, $\bar\lambda_{\dot\alpha i}$ are defined by
generating function lower.

The transformation~(\ref{x})--(\ref{l}) is canonical
transformation. The generating function of the canonical
transformation from system with phase variables $x^\mu$, $p_\mu$;
$\zeta^\alpha$, $\bar\zeta^{\dot\alpha}$, $p_{\zeta\alpha}$, $\bar
p_{\zeta\dot\alpha}$; $v_\alpha^i$, $\bar v_{\dot\alpha i}$,
$p_v{}^\alpha_{i}$, $\bar p_v{}^{\dot\alpha i}$ to the system with
phase variables $x^{(0)}$, $x_{(r)}$, $p_{(0)}$, $p_{(r)}$;
$\xi^i$, $\bar\xi_i$, $p_{\xi i}$, $\bar p_{\xi}^i$;
$\lambda_\alpha^i$, $\bar\lambda_{\dot\alpha i}$,
$\bar\omega^\alpha_i$, $\omega^{\dot\alpha i}$ has the form
\begin{eqnarray}
F &=& p_{(0)} x^{(0)}(x,v ,\bar v ) + p_{(r)} x_{(r)}(x,v ,\bar v)
+ p_{\xi i} \xi^i(\zeta ,v ) +
\bar p_{\xi}^i \bar\xi_i(\bar\zeta ,\bar v ) \nonumber\\
&&{} +\bar\omega^\alpha_i \lambda_\alpha^i (v)+
\bar\lambda_{\dot\alpha i}(\bar v)\omega^{\dot\alpha i} \, .
\label{gen-fun}
\end{eqnarray}
Here the expressions for new variables $x^{(0)}(x,v ,\bar v)$,
$x_{(r)}(x,v ,\bar v)$, $\xi^i(\zeta ,v)$, $\bar\xi_i(\bar\zeta
,\bar v)$, $\lambda_\alpha^i (v)$, $\bar\lambda_{\dot\alpha
i}(\bar v)$ in term of old variables from the right hand side of
the equations~(\ref{x}), (\ref{k}),  (\ref{l}) have been used.
That construction of the generating function reproduces the
expressions~(\ref{x})--(\ref{l}), whereas the expressions of
harmonic momenta $p_v{}^\alpha_{i}$, $\bar p_v{}^{\dot\alpha i}$
are
\begin{equation}\label{tran-om1}
p_v{}^\alpha_{i}=m^{1/2} \left[\bar\omega^{\alpha}_i + {\textstyle
\frac{1}{2m}}p_{(0)}\bar\lambda_{\dot\alpha i}
x^{\dot\alpha\alpha} + {\textstyle
\frac{1}{2m}}p_{(r)}\bar\lambda_{\dot\alpha j}
x^{\dot\alpha\alpha}(\sigma_r )_i{}^j + p_{\xi
i}\zeta^\alpha\right] \, ,
\end{equation}
\begin{equation}\label{tran-om2}
\bar p_v{}^{\dot\alpha i}=m^{1/2} \left[\omega^{\dot\alpha i} +
{\textstyle
\frac{1}{2m}}p_{(0)}x^{\dot\alpha\alpha}\lambda_{\alpha}^i +
{\textstyle \frac{1}{2m}}p_{(r)}
x^{\dot\alpha\alpha}\lambda_{\alpha}^j(\sigma_r )_j{}^i +\bar
p_{\xi}^i \bar\zeta^{\dot\alpha}\right] \, .
\end{equation}
Therefore in new variables the constraints~(\ref{const-Tr}),
(\ref{const-Vec}) acquire the additional terms
\begin{equation}\label{D0}
D_0 ={\cal D}_0 -i(\bar\xi_i\bar p_\xi^i -p_{\xi i}\xi^i) \approx
0 \, ,
\end{equation}
\begin{equation}\label{B0}
B_0 = {\cal B}_0 +2x^{(0)}p_{(0)} +2x_{(r)}p_{(r)} +(\bar\xi_i\bar
p_\xi^i +p_{\xi i}\xi^i)\approx 0 \, ,
\end{equation}
\begin{equation}\label{Dr}
D_r ={\cal D}_r - {\textstyle \frac{1}{2}}
\epsilon_{rsp}p_{(s)}x_{(p)} -{\textstyle \frac{i}{2}}(\sigma_r
)_i{}^j (\bar\xi_j\bar p_\xi^i -p_{\xi j}\xi^i) \approx 0 \, ,
\end{equation}
\begin{equation}\label{Br}
B_r = {\cal B}_r +x_{(r)}p_{(0)} +x^{(0)}p_{(r)} + {\textstyle \frac{1}{2}}(\sigma_r )_i{}^j (\bar\xi_j\bar p_\xi^i +p_{\xi
j}\xi^i) \approx 0\, ,
\end{equation}
where in new constraints $D_0$, $B_0$, $D_r$, $B_r$  the
expressions for ${\cal D}_0$, ${\cal B}_0$, ${\cal D}_r$, ${\cal
B}_r$ as in~(\ref{const-Tr}), (\ref{const-Vec}) with new
$\lambda_\alpha^i$, $\bar\lambda_{\dot\alpha i}$,
$\bar\omega^\alpha_i$, $\omega^{\dot\alpha i}$ in place of
$v_\alpha^i$, $\bar v_{\dot\alpha i}$, $p_v{}^\alpha_{i}$, $\bar
p_v{}^{\dot\alpha i}$
\begin{equation}\label{con-Tr}
{\cal D}_0 \equiv i(\bar\omega^{\alpha}_{i}\lambda_\alpha^i - \bar
\lambda_{\dot\alpha i}\omega^{\dot\alpha i})\approx 0 \, ,\quad
{\cal B}_0 \equiv \bar\omega^{\alpha}_{i}\lambda_\alpha^i + \bar
\lambda_{\dot\alpha i}\omega^{\dot\alpha i}\approx 0 \, ,
\end{equation}
\begin{equation}\label{con-Vec}
{\cal D}_r \equiv {\textstyle \frac{i}{2}}(\sigma_r )_i{}^j
(\bar\omega^{\alpha}_{j}\lambda_\alpha^i - \bar
\lambda_{\dot\alpha j}\omega^{\dot\alpha i})\approx 0 \, ,\quad
{\cal B}_r \equiv {\textstyle \frac{1}{2}}(\sigma_r )_i{}^j
(\bar\omega^{\alpha}_{j}\lambda_\alpha^i + \bar
\lambda_{\dot\alpha j}\omega^{\dot\alpha i})\approx 0 \, .
\end{equation}

The constraints~(\ref{s1})--(\ref{s3}) in new variables are
\begin{equation}\label{new-s1}
-(p_{(0)})^2 + p_{(r)}p_{(r)} +m^2\approx 0 \, ,
\end{equation}
\begin{equation}\label{new-s2}
-{\textstyle \frac{1}{m}}\xi^i p_i{}^j \bar\xi_j-j\approx 0 \, ,
\end{equation}
\begin{equation}\label{new-s3}
v^\alpha_i d_{\zeta\alpha}=m^{-1/2}\left(ip_{\xi i}+{\textstyle \frac{1}{m}} p_i{}^j \bar\xi_j\right) \approx 0\, ,\qquad \bar
d_{\zeta\dot\alpha} \bar v^{\dot\alpha i} = m^{-1/2}\left(i\bar p_{\xi}^i -{\textstyle \frac{1}{m}} \xi^j p_j{}^i \right)
\approx 0\, ,
\end{equation}
where $p_i{}^j \equiv p_{(0)}\delta_i^j +p_{(r)}(\sigma_r )_i{}^j$
(in term of initial space--time momentum $p_i{}^j=v^{\alpha}_i
p_{\alpha\dot\alpha}\bar v^{\dot\alpha j}
=\frac{1}{m}\lambda^{\alpha}_i
p_{\alpha\dot\alpha}\bar\lambda^{\dot\alpha j}$).

The transformations generated the constraints~(\ref{Br}) are
Wigner transformations. By these transformations we can transform
four--momentum (in harmonic basis) to standard form with
\begin{equation}\label{stand}
p_{(r)} \approx 0\, ,\quad r=1,2,3 \, .
\end{equation}
This conditions are gauge fixing conditions for constraints $B_r
\approx 0$ from which we obtain the expressions for $x_{(r)}$.
Because of resolved form of the gauge fixing
conditions~(\ref{stand}) the Poisson brackets for another
variables do not exchanged. Now the mass--shell
condition~(\ref{new-s1}) takes the form
\begin{equation}\label{p0}
p_{(0)} \pm m \approx 0\, ,
\end{equation}
fixed by means condition
\begin{equation}\label{x0}
x^{(0)} \approx 0\, ,
\end{equation}
which has resolved form also.

The constraints~(\ref{new-s2}), (\ref{new-s3}) take the form
\begin{equation}\label{n-s2}
\pm \bar\xi_i \xi^i - j\approx 0 \, ,
\end{equation}
\begin{equation}\label{n-s3}
\psi_i\equiv ip_{\xi i}\mp \bar\xi_i \approx 0\, ,\qquad
\bar\psi^i\equiv i\bar p_{\xi}^i \pm \xi^i \approx 0 \, .
\end{equation}
The last constraints (\ref{n-s3}) $\psi_i\approx 0$, $\bar\psi^i
\approx 0$ are the pairs of second class constraints, $\{\psi_i ,
\bar\psi^j\}=\mp 2i\delta^j_i$. After introducing Dirac brackets
for them
$$
\left\{A , B\right\}^* = \left\{A , B\right\} \mp {\textstyle
\frac{i}{2}}\left( \{A , \psi_i\} \{\bar\psi^i , B\} - \{A ,
\bar\psi^i\} \{\psi_i , B\} \right)
$$
the variables $p_{\xi i}$, $\bar p_{\xi}^i$ are excluded whereas
remaining variables $\xi^i$, $\bar\xi_i$ have nonzero Dirac
brackets $\{\xi^i , \bar\xi_j\}^*=\mp \frac{i}{2}\delta^i_j$. The
variables $\xi^i$ and $\bar\xi_i$ is canonically conjugate each
other. This construction generate automatically corresponding
kinetic term for $\xi^i$ and $\bar\xi_i$ in Lagrangian lower.

Thus, we exclude completely the space--time variables and obtain
the system with variables $\lambda_\alpha^i$,
$\bar\lambda_{\dot\alpha i}$, $\bar\omega^\alpha_i$,
$\omega^{\dot\alpha i}$, $\xi^i$, $\bar\xi_i$ and constraints
\begin{equation}\label{n-h}
h\equiv\lambda^{\alpha i}\lambda_{\alpha i} +2m\approx 0 \,
,\qquad \bar h\equiv\bar\lambda_{\dot\alpha i}
\bar\lambda^{\dot\alpha i} +2m\approx 0 \, ,
\end{equation}
\begin{equation}\label{l-D0}
D_0 ={\cal D}_0 \pm 2\bar\xi_i \xi^i =
i(\bar\omega^{\alpha}_{i}\lambda_\alpha^i -
\bar\lambda_{\dot\alpha i}\omega^{\dot\alpha i}) \pm 2\bar\xi_i
\xi^i \approx 0 \, ,
\end{equation}
\begin{equation}\label{l-Dr}
D_r ={\cal D}_r \pm (\sigma_r )_i{}^j \bar\xi_j \xi^i = (\sigma_r
)_i{}^j \left[ {\textstyle \frac{i}{2}}
(\bar\omega^{\alpha}_{j}\lambda_\alpha^i - \bar\lambda_{\dot\alpha
j}\omega^{\dot\alpha i}) \pm \bar\xi_j \xi^i \right] \approx 0 \,,
\end{equation}
\begin{equation}\label{l-s}
\emph{\textbf{S}}\equiv S-j \equiv\pm\bar\xi_i \xi^i - j\approx 0
\, ,
\end{equation}
\begin{equation}\label{l-B0}
B_0 = {\cal B}_0 = \bar\omega^{\alpha}_{i}\lambda_\alpha^i +
\bar\lambda_{\dot\alpha i}\omega^{\dot\alpha i}\approx 0 \, .
\end{equation}

The last constraint $B_0\approx 0$~(\ref{l-B0}) and constraint
$h+\bar h \approx 0$ form pair of self-conjugated second class
constraints. We can consider the constraint $B_0\approx 0$ as
gauge fixing condition for constraint $h+\bar h \approx 0$. Thus
we have equivalent system which has the
constraints~(\ref{n-h})-(\ref{l-s}). Of course we can impose the
constraint $B_0\approx 0$ in arbitrary moment.

Let us verify the equality of number of physical degrees of
freedom in twistor system with constraints~(\ref{n-h})-(\ref{l-s})
and massive spinning particle with Lagrangian~(\ref{lagr-s,m}).
The constraint~(\ref{l-s}) $S-j \approx 0$ and traceless
parts~(\ref{l-Dr}) $D_r \approx 0$ are first class constraints.
Also first class constraints is the constraint $h -\bar h \approx
0$, whereas constraint $h +\bar h \approx 0$ and trace
part~(\ref{l-D0}) $D_0 \approx 0$ are conjugate each other second
class constraints. Thus in twistor variables we have $5$ of first
class constraints and $2$ of second class ones. These constraints
exclude $12$ degrees of freedom. Since phase space, which contain
$\lambda_\alpha^i$, $\omega^{\dot\alpha i}$ and $\xi^i$, have $20$
variables, number of the physical degrees of freedom in twistor
model with constraints~(\ref{n-h})-(\ref{l-s}) are $8$. This
coincides with number of the physical degrees of freedom in
space--time formulation of massive spinning particle with
Lagrangian~(\ref{lagr-s,m}). Here we have $16$ variables in
$x^\mu$, $p_\mu$, $\zeta^\alpha$ and $p_{\zeta\alpha}$ and $2$ of
first class constraints (spin constraint and mass--shell
constraint) and $4$ of second class constraints (spinor
constraints). Thus the number of of physical degrees of freedom is
also $8$.

Note that constraints~(\ref{l-D0}) and (\ref{l-Dr}) are inscribed
in form of constraints
\begin{equation}\label{l-D}
D_i{}^j \equiv {\textstyle \frac{i}{2}}\left(
\bar\omega^{\alpha}_i \lambda_\alpha^j - \bar\lambda_{\dot\alpha
i} \omega^{\dot\alpha j} \right) \pm \bar\xi_i \xi^j \approx 0 \,
.
\end{equation}
Then $D_0 \approx 0$ is defined trace part of $D_i{}^j \approx 0$,
$D_0 =D_i{}^i$, whereas $D_r \approx 0$ are proportional traceless
parts $D_i{}^j -\frac{1}{2}\delta_i{}^j D_k{}^k\approx 0$, $D_r
=(\sigma_r )_i{}^j D_j{}^i$.
\par\bigskip
\centerline{\bf TWISTOR TRANSFORMATION } \centerline{\bf AND
LAGRANGIAN OF MASSIVE SPINNING PARTICLE IN TWISTOR FORMULATION}
\par
Expressions obtained in previous section give us directly full set
of equations defined twistor transformation.

Using completeness conditions for harmonics $v_\alpha^i$
(\ref{norm}) or for spinors $\lambda_\alpha^i$ (\ref{n-h}) we have
\begin{equation}
p_{\alpha\dot\alpha}=-{\textstyle \frac{1}{m}}\lambda_\alpha^i
p_i{}^j \bar\lambda_{\dot\alpha j}=-{\textstyle
\frac{1}{m}}\lambda_\alpha^i ( p_{(0)}\delta_i^j +p_{(r)}(\sigma_r
)_i{}^j)\bar\lambda_{\dot\alpha j}\, .
\end{equation}
Then after gauge fixing (adaptation of harmonic basis to
space--time one) (\ref{stand}) $p_{(r)} = 0$, $r=1,2,3$ and
mass--shall condition~(\ref{p0})  $p^{(0)} =-p_{(0)}= \pm m$ we
obtain twistor--like representation for four--momentum
\begin{equation}\label{tw-p}
p_{\alpha\dot\alpha}=\pm\lambda_\alpha^i \bar\lambda_{\dot\alpha
i}\, .
\end{equation}
From expressions (\ref{k}) and using (\ref{l}) we have expressions
for new variables $\xi^i$, $\bar\xi_i$ in term of spinning
variables $\lambda_\alpha^i$, $\bar \lambda_{\dot\alpha i}$
\begin{equation}\label{tw-x}
\xi^i =\zeta^\alpha \lambda_\alpha^i \, ,\qquad \bar\xi_i =\bar
\lambda_{\dot\alpha i}\bar\zeta^{\dot\alpha} \, .
\end{equation}
From canonical transformation (\ref{tran-om1}), (\ref{tran-om2}) and using conditions (\ref{stand}), (\ref{p0}) and
(\ref{const-om}), we obtain incidence conditions
\begin{equation}\label{tw-o}
\bar\omega^{\alpha}_i =\pm {\textstyle \frac{1}{2}}\bar\lambda_{\dot\alpha i} x^{\dot\alpha\alpha} \pm i\bar\xi_i\zeta^\alpha \,
, \qquad \omega^{\dot\alpha i} =\pm{\textstyle \frac{1}{2}}x^{\dot\alpha\alpha}\lambda_{\alpha}^i \mp i\xi^i
\bar\zeta^{\dot\alpha}\, ,
\end{equation}
which defined $\omega$--spinors by another variables. Thus we
obtain twistor transformations~(\ref{tw-p})-(\ref{tw-o}) of
massive spinning particle from space--time formulation in terms of
the variables $x^\mu$, $p_\mu$; $\zeta^\alpha$,
$\bar\zeta^{\dot\alpha}$ to twistor formulation in terms of the
variables $\lambda_\alpha^i$, $\bar\lambda_{\dot\alpha i}$,
$\bar\omega^\alpha_i$, $\omega^{\dot\alpha i}$; $\xi^i$,
$\bar\xi_i$. Let us summarize the obtained twistor transition.

In twistor formulation of massive particle we use two spinors
\begin{equation}
\lambda_\alpha^i \, ,\quad \bar\lambda_{\dot\alpha i} =
\overline{(\lambda_\alpha^i)}  \, ,\quad i=1,2
\end{equation}
in form of them the four--momentum of particle has the resolved
form~(\ref{tw-p}). For fulfilment of the mass--shell condition
$$
p_{\alpha \dot\alpha}p^{\dot\alpha\alpha}-2m^2 = -2(p^2 +m^2)
\approx 0 \,
$$
the spinors $\lambda_\alpha^i$ are subjected to the
conditions~(\ref{n-h})
\begin{equation}\label{norm}
\lambda^{\alpha i}\lambda_{\alpha i} +2m\approx 0 \, ,\quad \bar\lambda_{\dot\alpha i} \bar\lambda^{\dot\alpha i} +2m\approx 0\,
,
\end{equation}
where $\lambda_{\alpha i} =\epsilon_{ij}\lambda_\alpha^j$,
$\bar\lambda_{\dot\alpha}^i =\epsilon^{ij}\bar\lambda_{\dot\alpha
j}$ and components of skew--symmetric tensor $\epsilon^{ij}$ are
equal matrix elements of matrix $i\sigma_2$,
$\epsilon^{ij}\epsilon_{jk}=\delta^i_k$. The conditions can be
inscribed in the equivalent form
\begin{equation}
\lambda^{\alpha i}\lambda_{\alpha}^j -m\epsilon^{ij}\approx 0 \, ,\quad
\bar\lambda_{\dot\alpha i} \bar\lambda^{\dot\alpha}_j
-m\epsilon_{ij}\approx 0
\end{equation}
or also in form
\begin{equation}
\lambda^{\alpha i}\lambda^{\beta}_i -m\epsilon^{\alpha\beta}\approx 0
\, ,\quad
\bar\lambda_{\dot\alpha i} \bar\lambda_{\dot\beta}^i
-m\epsilon_{\dot\alpha\dot\beta}\approx 0 \, .
\end{equation}

For each spinor $\lambda_\alpha^i$, $i=1,2$ it is canonically
conjugated spinor
\begin{equation}\label{omega}
\omega^{\dot\alpha i}\, ,\quad \bar\omega^{\alpha}_i =\overline{(\omega^{\dot\alpha i})}\, ,\quad i=1,2\, ,
\end{equation}
which play the role of the second spinor component of
corresponding twistor. The incidence conditions which defined
$\omega$--spinors by another variables have the form~(\ref{tw-o}).
After contractions with $\bar\lambda_{\dot\alpha i}$ and
$\lambda_\alpha^i$ the incidence conditions~(\ref{tw-o}) give us
the following constraints~(\ref{l-D}) (or, equivalently,
(\ref{l-D0}) and (\ref{l-Dr})).

In twistor formulation, particle spin is described by means of two
complex scalar variables
\begin{equation}\label{bexi}
\xi^i \, ,\quad \bar\xi_i =\overline{(\xi^i)} \, ,\quad
i=1,2 \, .
\end{equation}
Connection of them with index spinor $\zeta$ (spinning variables
in space--time formulation) is defined by
expressions~(\ref{tw-x}). The variables~(\ref{bexi}) satisfy the
constraint~(\ref{l-s}).

Such form of twistor transformations~(\ref{tw-p})-(\ref{tw-o})
give us desired form of kinetic terms in twistor variables.
Precisely, using~(\ref{tw-p})-(\ref{tw-o}) the kinetic terms in
twistor variables are
\begin{equation}\label{kin-term-be}
- {\textstyle
\frac{1}{2}}p_{\alpha\dot\alpha}\Pi^{\dot\alpha\alpha} = \mp
{\textstyle \frac{1}{2}}(d\bar\omega^\alpha_i \,\lambda_\alpha^i -
d\bar\lambda_{\dot\alpha i} \,\omega^{\dot\alpha i} -
\bar\omega^\alpha_i \,d\lambda_\alpha^i + \bar\lambda_{\dot\alpha
i} \,d\omega^{\dot\alpha i}) \pm i(d\bar\xi_i\, \xi^i -\bar\xi_i\,
d\xi^i )\, .
\end{equation}

Spinors $\lambda^i$ and $\omega^i$ are combined in twistors $Z^i$
by
\begin{equation}\label{betwist}
Z_a^i =\left(\lambda_\alpha^i\, ,\,\,\, \omega^{\dot\alpha
i}\right) \, .
\end{equation}
If we introduce in standard way conjugate twistors for
ones~(\ref{betwist}) by
\begin{equation}\label{twist-conj-be}
\bar Z_{\dot a i} =\overline{(Z_a^i)} = \left(\bar\lambda_{\dot\alpha i}\, ,\,\,\, \bar\omega^{\alpha}_i \right) \, ,\quad \bar
Z^{a} = g^{a\dot b} \bar Z_{\dot b} = \left(\bar\omega^{\alpha}_i\, ,\,\,\, -\bar\lambda_{\dot\alpha i} \right)\, ,
\end{equation}
where
$$
g^{a\dot b}=\left(
\begin{array}{cc}
0 & \delta^{\alpha}{}_{\beta} \\
-\delta_{\dot\alpha}{}^{\dot\beta} & 0
\end{array}
\right)
$$
then the kinetic terms~(\ref{kin-term-be}) can be rewritten as
\begin{equation}\label{kin-term-2}
p_\mu \Pi^\mu=-{\textstyle
\frac{1}{2}}p_{\alpha\dot\alpha}\Pi^{\dot\alpha\alpha} = \mp
{\textstyle \frac{1}{2}}(d\bar Z^a_i\, Z_a^i -\bar Z^a_i\, dZ_a^i)
\pm i(d\bar\xi_i\, \xi^i -\bar\xi_i\, d\xi^i )\, .
\end{equation}

Thus the twistor formulation of massive spinning particle is
described by Lagrangian
\begin{eqnarray}
L &=& \mp{\textstyle \frac{1}{2}}\left(\dot{\bar Z}^a_i\, Z_a^i -
\bar Z^a_i\, \dot{Z}_a^i \right)
-i\left(\dot{\bar\xi}_i\, \xi^i -\bar\xi_i\, \dot{\xi}^i \right) \nonumber\\
&& {}- M\left( S-j \right)-N_j{}^i D_i{}^j- K h -\bar K\bar h\, , \label{lagr-betwist}
\end{eqnarray}
where $M$, $N_j{}^i$, $K$ and $\bar K$ are Lagrange multipliers
for constraints~(\ref{l-s}), (\ref{l-D}) and (\ref{n-h}).

It is note that twistor mass--shell constraints~(\ref{n-h}) can be
rewritten in form
\begin{equation}\label{norm-2}
h= Z_a^i I^{ab} Z_{bi} +2m \approx 0 \, ,\quad \bar h = \bar Z^a_i I_{ab} \bar Z^{bi} +2m \approx 0\, ,
\end{equation}
if we use so--called infinity twistors (asymptotic twistors)
$$
I^{ab}=\left(
\begin{array}{cc}
\epsilon^{\alpha\beta} & 0 \\
0 & 0
\end{array}
\right) \, , \quad
I_{ab}=\left(
\begin{array}{cc}
0 & 0 \\
0 & \epsilon^{\dot\alpha\dot\beta}
\end{array}
\right) \, .
$$
Also the constraints~(\ref{l-D}) are represented in covariant
contractions of twistors
\begin{equation}\label{const-u2-cov}
D_i{}^j = {\textstyle \frac{i}{2}} \bar Z^a_i Z_a^j \pm \bar\xi_i
\xi^j \approx 0 \, .
\end{equation}
We can introduce also so--called `bosonic supertwistors'
\begin{equation}\label{G-twist-be}
{\cal Z}_{\cal A}^i =(Z_a^i ;\xi^i ) \, ,\quad \bar{\cal Z}^{{\cal
A}}_i = (\bar Z^{a}_i ;\mp 2i\bar\xi_i )
\end{equation}
in form of them the kinetic terms of Lagrangian~(\ref{lagr-betwist}) are
rewritten as
\begin{equation}\label{Lagr-be-sup}
\mp {\textstyle \frac{1}{2}} \left( \dot{\bar{\cal Z}}^{{\cal
A}}_i {\cal Z}_{{\cal A}}^i - \bar{\cal Z}^{{\cal A}}_i \dot{\cal
Z}_{{\cal A}}^i \right)
\end{equation}
and the constraints~(\ref{const-u2-cov}) are
\begin{equation}\label{u2-sup}
D_i{}^j = {\textstyle \frac{i}{2}} \bar{\cal Z}^{{\cal A}}_i {\cal
Z}_{{\cal A}}^j \approx 0 \, .
\end{equation}
\par\bigskip
\centerline{\bf QUANTIZATION OF THE TWISTORIAL SPINNING PARTICLE}
\par
Let us carry out canonical quantization a la Dirac of massive spinning particle in twistor formulation. For definiteness, we
consider case with upper sign in Lagrangian~(\ref{lagr-betwist}). The system is described by twistor variables
$\bar\omega^{\alpha}_i$, $\omega^{\dot\alpha i}$, $\lambda_\alpha^i$, $\bar\lambda_{\dot\alpha i}$, $\xi^i$, $\bar\xi_i$ and
constraints~(\ref{n-h})--(\ref{l-s}), i.e.
\begin{equation}\label{spi-tw}
S-j \equiv \bar\xi_i \xi^i -j\approx 0 \, ,
\end{equation}
$SU(2)$--constraints
\begin{equation}\label{con-su2}
D_r =(\sigma_r )_i{}^j D_j{}^i ={\textstyle \frac{i}{2}} \left[
\lambda_\alpha^i (\sigma_r )_i{}^j \bar\omega^{\alpha}_{j}-
\omega^{\dot\alpha i}) (\sigma_r )_i{}^j \bar\lambda_{\dot\alpha
j} \right] + \xi^i (\sigma_r )_i{}^j \bar\xi_j \approx 0 \, ,
\end{equation}
the normalization conditions
\begin{equation}\label{nor}
h\equiv\lambda^{\alpha i}\lambda_{\alpha i} +2m\approx 0 \, ,\quad
\bar h\equiv\bar\lambda_{\dot\alpha i} \bar\lambda^{\dot\alpha i}
+2m\approx 0
\end{equation}
and $U(1)$--constraint
\begin{equation}\label{u1}
D_0 = 2D_i{}^i =
i(\lambda_\alpha^i\bar\omega^{\alpha}_{i} -
\omega^{\dot\alpha i}\bar\lambda_{\dot\alpha i}) + 2\xi^i\bar\xi_i
\approx 0 \, .
\end{equation}
The constraint $D_0$ is gauge fixing condition for constraint
$h-\bar h\approx 0$. We impose also gauge fixing
condition~(\ref{l-B0})
\begin{equation}\label{gfb0}
B_0 = \lambda_\alpha^i  \bar\omega^{\alpha}_{i} +
\omega^{\dot\alpha i} \bar\lambda_{\dot\alpha i} \approx 0
\end{equation}
for constraint $h+\bar h\approx 0$ and regard that the
constraint~(\ref{nor})--(\ref{gfb0}) are fulfilled in strong
sense. Introduction of Dirac brackets for them do not change
commutation relations of another constraints~[23], [24]  and we
may consider the twistor variables $\omega$ and $\lambda$ with
standard canonical relations
\begin{equation}\label{com-rel-lo}
\{\lambda_\alpha^i , \bar\omega^{\beta}_j\}= \delta^\beta_\alpha
\delta_j^i \, , \quad \{\bar\lambda_{\dot\alpha i},
 \omega^{\dot\beta j}\}=
\delta^{\dot\beta}_{\dot\alpha}\delta^j_i
\end{equation}
in $SU(2)$--constraints~(\ref{con-su2}). The kinetic terms
in~(\ref{lagr-betwist}) lead to following Dirac brackets for
$\xi$--variables
\begin{equation}\label{com-rel-xi}
\{ \bar\xi_i, \xi^j \}^{*}=- {\textstyle \frac{i}{2}}\delta^j_i \,
.
\end{equation}

We use the realization of spinor variables, quantum commutators of which are
$$
[ \bar\omega^{\alpha}_i, \lambda_\beta^j ]=- i\delta^\alpha_\beta
\delta_i^j \, , \quad [ \omega^{\dot\alpha i},
\bar\lambda_{\dot\beta j} ]=-
i\delta^{\dot\alpha}_{\dot\beta}\delta^i_j \, ,
$$
as differential operators. For definiteness we take representation
with diagonal $\lambda$--spinors whereas realization of
$\omega$--spinors in constraints~(\ref{con-su2}) is
$$
\bar\omega^{\alpha}_i =-i {\partial}/{\partial\lambda_\alpha^i}
\,,\quad \omega^{\dot\alpha i} =-i
{\partial}/{\partial\bar\lambda_{\dot\alpha i}}\,.
$$
Quantum algebra of $\xi$--variables are
$$
[ \xi^i , \bar\xi_j ]= -{\textstyle \frac{1}{2}}\delta^i_j \, .
$$
The variables
\begin{equation}\label{a}
a_i \equiv \sqrt{2} \bar\xi_i \, ,\quad a^{+i} \equiv \sqrt{2}
\xi^i
\end{equation}
are usual annihilation and creation operators of two--dimensional
oscillator
$$
\left[ a_i , a^{+j} \right]= \delta^j_i \, .
$$
The wave function
will be taken in filling numbers space of these operators.

Thus the wave function $\Psi(\lambda ,\bar\lambda )$ is subjected the
first class constraints
\begin{equation}\label{spin-quant}
\left( S-J \right)\Psi \equiv \left({\textstyle \frac{1}{2}}a^{+i}
a_i -J \right)\Psi =0 \, ,
\end{equation}
\begin{equation}\label{su2-quant}
D_r \Psi =\left( {\cal D}_r +\Delta_r \right) \Psi =0 \,,\quad r=1,2,3\, ,
\end{equation}
where
\begin{equation}\label{Dr-quant}
{\cal D}_r \equiv {\textstyle \frac{1}{2}}\left[ \lambda_\alpha^i
(\sigma_r )_i{}^j \frac{\partial}{\partial\lambda_\alpha^j}-
\frac{\partial}{\partial\bar\lambda_{\dot\alpha i}} (\sigma_r
)_i{}^j \bar\lambda_{\dot\alpha j} \right] \, ,
\end{equation}
\begin{equation}\label{Delta-quant}
\Delta_r \equiv {\textstyle \frac{1}{2}} a^{+i} (\sigma_r )_i{}^j
a_j \,.
\end{equation}
The constant $J$ in constraint~(\ref{spin-quant}) is classical
constant $j$ in~(\ref{spi-tw}) renormalized ordering constants.

The operators ${\cal D}_r$ and $\Delta_r$ form $SU(2)$--algebras
$$
\left[ {\cal D}_r ,{\cal D}_s \right]=i\epsilon_{rsp}{\cal D}_p \,
,\quad \left[ \Delta_r , \Delta_s \right]=i\epsilon_{rsp} \Delta_p
\, .
$$
The possible ordering constants in operators ${\cal D}_3$ and $\Delta_3$
mutually compensate each other. Otherwise the quantum algebra of
first class constraints $D_r$
$$
\left[ D_r , D_s \right]=i\epsilon_{rsp} D_p
$$
will not be closed.
\par\bigskip
\centerline{\bf ANALYSIS OF SPECTRUM}
\par
Let us find the possible values of spin in spectrum. The direct
method for finding of particle spin in spectrum is determination
of eigenvalues of Casimir operators of Poincare group. The
operator of the four--translations in realization on the space of
wave function $\Psi(\lambda ,\bar\lambda )$ has the form
\begin{equation}\label{4-trans}
P_{\alpha\dot\alpha}=\lambda_\alpha^i \bar\lambda_{\dot\alpha i}
\end{equation}
whereas the operator of Lorentz transformations is
\begin{equation}\label{Lor-trans}
M_{\alpha\dot\alpha\beta\dot\beta}= 2i ( \epsilon_{\dot\alpha\dot\beta} M_{\alpha\beta} + \epsilon_{\alpha\beta} \bar
M_{\dot\alpha\dot\beta} )\,,
\end{equation}
where
$$
M_{\alpha\beta} =\lambda_{(\alpha}^i
\frac{\partial}{\partial\lambda^{\beta )i}} \, ,\quad \bar
M_{\dot\alpha\dot\beta} = \bar\lambda_{(\dot\alpha i}
\frac{\partial}{\partial\bar\lambda^{\dot\beta )i}} \, .
$$

In consequence of normalization conditions~(\ref{nor}) of
$\lambda$--spinors we have on physical states
\begin{equation}\label{mas}
P^2 = -m^2
\end{equation}
i.e. the physical states describe the particle of mass $m$.

By means direct calculations we obtain that Pauli--Lubanski
pseudovector
$$
W_{\alpha\dot\alpha} =P_{\alpha}^{\dot\beta}\bar
M_{\dot\beta\dot\alpha} -P_{\dot\alpha}^\beta M_{\beta\alpha}
$$
take the form
\begin{equation}\label{PL}
W_{\alpha\dot\alpha} = iu_{r{}\alpha\dot\alpha} {\cal D}_r\, ,
\end{equation}
where
$$
u_{r{}\alpha\dot\alpha} \equiv
\lambda_\alpha^i (\sigma_r)_i{}^j \bar\lambda_{\dot\alpha j} \, ,\quad
u_r \cdot u_s =-m^2\delta_{rs}
$$
and operators ${\cal D}_r$ are the same as in~(\ref{Dr-quant}).
Note that $[ D_r , u_{s{}\alpha\dot\alpha} ]= \epsilon_{rsp}
u_{p{}\alpha\dot\alpha}$. Right now we obtain
\begin{equation}\label{Casim}
W^2 =m^2 {\cal D}_r {\cal D}_r \, .
\end{equation}

But from~(\ref{su2-quant}) we see that on physical states
${\cal D}_r = D_r - \Delta_r$ and ${\cal D}_r {\cal D}_r =
D_r D_r -2\Delta_r D_r +\Delta_r \Delta_r$, i. e. on states of spectrum
\begin{equation}\label{Casim1}
W^2 =m^2 \Delta_r \Delta_r \, .
\end{equation}
But direct calculation gives us that
$$
\Delta_r \Delta_r = {\textstyle \frac{1}{2}}a^{+i} a_i (
{\textstyle \frac{1}{2}}a^{+i} a_i +1 ) =S ( S +1 ) \, .
$$
In consequence of constraint~(\ref{spin-quant}) the operator $S$ is equal
$J$ on physical states. Therefore on states of spectrum
\begin{equation}\label{Casim2}
W^2 =m^2 S ( S +1 )
\end{equation}
i.e. in spectrum we have massive particle with fixed spin which equal $J$.
\par\bigskip
\centerline{\bf WAVE FUNCTION OF TWISTORIAL MASSIVE PARTICLE}
\par
The operators $\Delta_r$ form $SU(2)$--algebra which realized by operators of two oscillators. Let integer non--negative numbers
$n_1$ and $n_2$ are corresponding filling numbers i.e. $n_1$ and $n_2$ are the eigenvalues of operators $a^{+1} a_1$ and $a^{+2}
a_2$. The constraints~(\ref{spin-quant}) gives us that $\frac{1}{2}(n_1 +n_2) =J\ge 0$. Then the number $\frac{1}{2}(n_1 -n_2)
\equiv M$ takes $(2J+1)$ values $M=-J,-J+1,...,J-1,J$. In normalized basis the action of operators $\Delta_\pm =\Delta_1 \pm
i\Delta_2$, $\Delta_3$ on wave function $\Psi (\lambda ,\bar\lambda )$, which has index $M$,
$$
\Psi_M (\lambda ,\bar\lambda )\,,\qquad M=-J,-J+1,...,J-1,J\,,
$$
is
$$
\Delta_3 \Psi_M =M\Psi_M \, ,\qquad \Delta_\pm \Psi_M=\sqrt{(J\mp
M)(J\pm M+1)} \Psi_{M\pm 1}\, .
$$
Then the action of constraints~(\ref{su2-quant}) on $(2J+1)$--component
wave function $\Psi_M (\lambda ,\bar\lambda )$ takes the form
\begin{equation}\label{eq}
{\cal D}_3 \Psi_M =-M\Psi_M \, ,\qquad {\cal D}_\pm \Psi_M=-\sqrt{(J\mp M)(J\pm M+1)} \Psi_{M\pm 1}\, ,
\end{equation}
where ${\cal D}_\pm ={\cal D}_1 \pm i{\cal D}_2$. All $(2J+1)$
components of wave function are obtained from one component, for
example from component of highest weight $\Psi_{+ J}$ or lowest
one $\Psi_{- J}$
$$
\Psi_M = \sqrt{\frac{(J\pm M)!}{(J\mp M)!(2J)!}} (-1)^M ({\cal
D}_\mp )^{J\mp M} \Psi_{\pm J} \, .
$$
These components $\Psi_{\pm J}$ are defined by equations
\begin{equation}\label{Psi}
{\cal D}_3 \Psi_{\pm J} =\pm J\Psi_{\pm J} \, ,\quad {\cal D}_\pm
\Psi_{\pm J} =0 \, ,\quad ({\cal D}_\mp )^{2J+1} \Psi_{\pm J} =0
\,.
\end{equation}

The operators ${\cal D}_r$ are generators of
$SU(2)$--transformations, acting on indices $i,j,k,...$ of
$SL(2,C)$--matrix $\lambda^i_\alpha$. The constraints~(\ref{eq})
state that the wave function $\Psi_M (\lambda ,\bar\lambda )$ is
defined up to local transformations acting on index $M$
\begin{equation} \label{su2-trans}
\Psi_M^\prime (\lambda^\prime) ={\bf D}^J_{MN}(h) \Psi_N (\lambda)\, ,
\end{equation}
where $h \in SU(2)$ and $\lambda^{\prime i}
_\alpha=h^i_j\lambda^j_\alpha$. The ${\bf D}^J_{MN}$ is matrix of
$SU(2)$--transformations of weight $J$. Thus the wave function is
defined in fact on homogeneous space ${\cal M} =G/H
=SL(2,C)/SU(2)$.

The harmonic expansion of function defined on $SL(2,C)$ is~[25]
\begin{eqnarray}
\Phi (\lambda ) &=& -\frac{1}{8\pi^2}\int {\rm Tr}
\left( F(\chi)T^{-1}_\chi (\lambda )\right) c(\chi)d\chi \nonumber\\
&=& -\frac{1}{32\pi^2}\int\limits_{-\infty}^\infty \sum_{n=-\infty}^\infty d\rho (n^2 +\rho^2){\rm Tr} \left( F(\chi)T^{-1}_\chi
(\lambda )\right)\, , \label{Fouir}
\end{eqnarray}
where Fourier transformation $F(\chi )$ acts on space of function
$\varphi (z)$, defined on two--dimensional complex plane with coordinates
$z=z^\alpha$, $\alpha =1,2$, by means of
$$
F(\chi) \varphi (z) \equiv \int \Phi (\lambda ) T_\chi (\lambda )
\varphi (z) d\lambda
$$
and $T_\chi$ is operator $SL(2,C)$--transformations
$$
T_\chi \varphi (z) =\varphi (z\lambda ) \, .
$$
In decomposition~(\ref{Fouir}) it is taken only
rep\-re\-sen\-ta\-ti\-ons of ba\-sic se\-ries $\chi =( (n+i\rho
)/2 , (-n+i\rho )/2)$, $c(\chi) =n^2 +\rho^2$.

For $SU(2)$--covariant function~(\ref{su2-trans})
$$
n=M \, .
$$
Therefore wave function of massive particle of spin $J$ has
harmonic decomposition on basic series of following form
\begin{equation} \label{harm-decomp}
\Psi_M (\lambda) =-\frac{1}{32\pi^2}\int\limits_{-\infty}^\infty d\rho (M^2 +\rho^2){\rm Tr} \left( F_M (\chi)T^{-1}_\chi
(\lambda )\right)\, ,
\end{equation}
where
$$
\chi =( (M+i\rho )/2 , (-M+i\rho )/2) \, .
$$

Thus as result quantization of the massive twistorial particle
with Lagrangian~(\ref{lagr-betwist}) we obtain in spectrum the
particle with fixed mass and fixed spin. The wave function of it
is defined by equations~(\ref{Psi}).
\par\bigskip
\centerline{\bf CONCLUSION}
\par
In this work we presented the twistor formulation of massive particle with arbitrary spin. This formulation is obtained from
massive spinning particle in index spinor formulation by means introducing pure gauge harmonic variables. After partial fixing
of gauges we obtain the model described two twistors (bitwistor) and two complex scalars. As result of canonical transformation
we obtain the conditions of twistor transformation. It is carried out quantization of the twistorial spinning particle. On
physical states Casimir operators of Poincare group have the value corresponding to the massive particle of fixing nonzero spin.
The wave function of twistorial massive particle is defined on homogeneous space $SL(2,C)/SU(2)$ and has harmonic expansion in
representations of basic series with one fixing weight.
\par\bigskip
This work was supported in part by INTAS Grant INTAS-2000-254 and by Ukrainian National Found of Fundamental Researches under
the Project \textsl{N} 02.07/383. We would like to thank I.A.~Bandos, A.~Frydryzhak, E.A.~Ivanov, S.O.~Krivonos, J.~Lukierski,
A.J.~Nurmagambetov and D.P.~Sorokin for interest to the work and for many useful discussion.
\par\bigskip
\centerline{\bf REFERENCES}
\par
{\small \begin{enumerate} \itemindent=-4mm  \item R. Penrose, J. Math. Phys. {\bf 8}(1967)345.\\
R. Penrose and M.A.H. MacCallum, Phys. Reports {\bf 6C}(1972)241.\\
 R. Penrose and W. Rindler, Spinors and Space-Time,
1986 (Cambridge: Cambridge University Press). \item L.P. Hughston, Twistor and Particles, Lecture Notes in Physics, Berlin, {\bf
97}, 1979, 153~pp. \item S.A. Hugget and K.P. Tod, An Introduction to the Twistor Theory, Camgrige University Press, 1994,
178~pp. \item Z. Perj\'{e}s, Phys. Rev. {\bf D 11}(1975)2031; Reports Math. Phys. {\bf 12}(1977)193; Phys. Rev. {\bf D
20}(1979)1857. \item E. Witten, Nucl. Phys. {\bf B 266}(1986)245. \item A.K.H. Bengtsson, I. Bengtsson, M. Cederwall and N.
Linden, Phys. Rev.
{\bf D 36}(1987)1766. \\
I. Bengtsson and M. Cederwall, Nucl. Phys. {\bf B 302}(1988)81. \item A. Ferber, Nucl. Phys. {\bf B 132}(1978)55. \item A.
Bette, J. Math. Phys. {\bf 25}(1984)2456, {\bf 37}(1996)1724. \item Y. Eisenberg and S. Solomon, Nucl. Phys. {\bf B
309}(1988)709; Phys. Lett. {\bf B 220}(1989)562. \item Y. Eisenberg, Phys. Lett. {\bf B 225}(1989)95. \item M.S. Plyshchay, Mod.
Phys. Lett. {\bf A 4}(1989)1827. \item D.P. Sorokin, V.I. Tkach
and D.V. Volkov, Mod. Phys. Lett. {\bf A4}(1989)901. \\
D.P. Sorokin, V.I. Tkach, D.V. Volkov and A.A. Zheltukhin, Phys. Lett. {\bf B 216}(1989)302. \item D.V. Volkov and A.A.
Zheltukhin, Lett. Math. Phys. {\bf 17}(1989)141; Nucl. Phys. {\bf B 335}(1990)723. \item V.A. Soroka, D.P. Sorokin, V.I. Tkach
and D.V. Volkov, On a twistor shift in particle and string dynamics, Preprint ITP UWR 765/90, Wroclaw, 1990, 9 pp.; JETP Lett.
{\bf 52}(1990)1124. \item P. Townsend, Phys. Lett. {\bf B 261}(1991)65. \item S. Fedoruk and V.G. Zima, Nucl. Phys. (Proc.
Suppl.) {\bf B 102\&103}(2001)233. \item V.G. Zima and S. Fedoruk, JETP Lett. {\bf 61}(1995)251. \item V.G. Zima and S. Fedoruk,
Class. Quantum Grav. {\bf 16}(1999)3653. \item V.G. Zima and S. Fedoruk, Phys. of Atomic Nucl. {\bf 63}(2000)617. \item S.
Fedoruk and V.G. Zima, Mod. Phys. Lett. {\bf A 15}(2000)2281. \item J. Wess and J. Bagger, Supersymmetry and Supergravity, 1983
(Princeton: Princeton University Press). \item A. Galperin, E. Ivanov, E. Kalizin, V. Ogievetsky and E. Sokatchev, Class.
Quantum Grav. {\bf 1}(1984)469; {\bf 2}(1985)155. \item
I.A. Bandos, Sov. J. Nucl. Phys. {\bf 51}(1990)906. \\
I.A. Bandos and A.A. Zheltukhin, Class. Quantum Grav. {\bf 12}(1995)609. \item V.G. Zima and S. Fedoruk, Theor. Math. Phys. {\bf
102}(1995)305. \item I.M. Gel'fand, M.I. Graev and N.Ya. Vilenkin, Generalized Functions, Vol.5, Integral Geometry and
Representation Theory, Academic Press, New York (1966).
\end{enumerate}}
\end{document}